# A Moran coefficient-based mixed effects approach to investigate spatially varying relationships

Daisuke Murakami[1,*], Takahiro Yoshida[1,2], Hajime Seya[3], Daniel A. Griffith[4], and Yoshiki Yamagata[1]

[1]Center for Global Environmental Research, National Institute for Environmental Studies,
16-2 Onogawa, Tsukuba, Ibaraki, 305-8506, Japan
Email: murakami.daisuke@nies.go.jp; yamagata@nies.go.jp

[2]Graduate School of Systems and Information Engineering, University of Tsukuba,
1-1-1 Tennodai, Tsukuba, Ibaraki, 305-8573, Japan
Email: yoshida.takahiro@sk.tsukuba.ac.jp

[3] Department of Civil Engineering, Graduate School of Engineering, Kobe University,
1-1 Rokkodai, Nada, Kobe, 657-8501, Japan
Email: hseya@people.kobe-u.ac.jp

[4]Schoold of Economic, Political and Policy Science, The University of Texas, Dallas,
800 W Campbell Rd, Richardson, TX, 75080, USA
Email: dagriffith@utdallas.edu

*: Corresponding author

**Abstract**

This study develops a spatially varying coefficient model by extending the random effects eigenvector spatial filtering model. The developed model has the following properties: its coefficients are interpretable in terms of the Moran coefficient; each of its coefficients can have a different degree of spatial smoothness; and it yields a variant of a Bayesian spatially varying coefficient model. Also, parameter estimation of the model can be executed with a relatively small computationally burden. Results of a Monte Carlo simulation reveal that our model outperforms a conventional eigenvector spatial filtering (ESF) model and geographically weighted regression (GWR) models in terms of the accuracy of the coefficient estimates and computational time. We empirically apply our model to the hedonic land price analysis of flood risk in Japan.

## 1. Introduction

Spatial heterogeneity is one of the important characteristics of spatial data (Anselin, 1988). Geographically weighted regression (GWR) (Fotheringham et al., 2002; Wheeler and Páez, 2009; Fotheringham, 2016) models are one useful approach for explicitly accounting for spatial heterogeneity of the model structure through spatially varying coefficients (SVCs). Thus far, GWR models have attained remarkable success with various applications, including social economic studies (e.g., Bitter et al., 2007; Huang et al., 2010), ecological studies (e.g., Wang et al., 2005; Austin, 2007), and health studies (e.g., Nakaya et al., 2005; Hu et al., 2012).

Despite the wide-ranging set of applications, existing studies have shown that the basic (original) GWR specification has several drawbacks. First, the coefficients of the basic GWR model typically suffer from multicollinearity (Páez et al., 2011; Wheeler and Tiefelsdorf, 2005). Second, the basic GWR model assumes the same degree of spatial smoothness for each coefficient, which is a rather strong assumption that fails to hold in most empirical applications. Fortunately, several extended GWR models have been proposed to address these problems. With regard to the first problem, Wheeler (2007; 2009) proposes regularized GWR models, by combining ridge and/or lasso regression with GWR, and its robustness in terms of the multicollinearity problem has been demonstrated. The limitations of regularized GWR models are its bias in coefficient estimates, just like conventional ridge and/or lasso regression. With regard to the second problem concerning uniform smoothers, Yang et al. (2014) and Lu et al. (2015) attempted to overcome this limitation.

The Bayesian spatially varying coefficients (B-SVC) model, based on a geostatistical (Gelfand et al., 2003) or lattice autoregressive approach (Assunçao, 2003), is another forms of the spatially varying coefficients models that requires Markov chain Monte Carlo (MCMC). Wheeler and Calder (2007) and Wheeler and Waller (2009) suggest that the coefficient estimates for the B-SVC model of Gelfand et al. (2003) are robust in terms of multicollinearity. In contrast to the GWR model, the B-SVC model allows differential spatial smoothness across coefficients. However, this differential makes computational costs prohibitive if a sample size is moderate to large (Finley, 2011). Integrated nested Laplace approximation (INLA)[1] based alternatives are becoming available now (Blangiardo and Cameletti, 2015; Congdon, 2014), but with unknown properties.

Hence, a SVC model with the following properties still needs to be developed: (a) robust to multicollinearity; (b) the possibility for each coefficient to have a different degree of spatial smoothness; and, (c) computational efficiency. This study develops a model satisfying these requirements by combining an eigenvector spatial filtering (ESF; Griffith 2003; Chun and Griffith, 2014) based SVC model (Griffith, 2008) and a random effects ESF (RE-ESF: Murakami and Griffith, 2015) model.

The following sections are organized as follows. Sections 2 and 3 introduce the GWR model and ESF-based SVC model of Griffith (2008), respectively. Section 4 introduces the RE-ESF model, and extends it to a SVC model. Section 5 compares the properties of our model with those of other SVC models. Section 6 summarizes results from a comparative Monte Carlo simulation experiment, and section 7 uses our model in a flood risk analysis. Section 8 concludes our discussion.

[1] See Rue et al.(2009) for the details about the INLA approach.

## 2. GWR models

The basic GWR model for a site $s_i \in D \subset \Re^2$ is formulated as follows:

$$y(s_i) = \sum_{k=1}^{K} x_k(s_i)\beta_k(s_i) + u(s_i)\,, \tag{1}$$

where $y(s_i)$ is the explained variable at location $s_i$, $x_k(s_i)$ is the $k$-th explanatory variable, $\beta_k(s_i)$ is the corresponding geographically varying coefficient, and $u(s_i)$ is a disturbance term. Estimates for the $\beta_k(s_i)$ coefficients are obtained with weighted least squares (WLS), which assigns greater weights to nearby observations than distant ones. The WLS estimator of $\boldsymbol{\beta}(s_i) = [\beta_1(s_i),\ldots \beta_K(s_i)]'$ is given by:

$$\hat{\boldsymbol{\beta}}(s_i) = [\mathbf{X}'\mathbf{G}(s_i)\mathbf{X})]^{-1}\mathbf{X}'\mathbf{G}(s_i)\mathbf{y}\,, \tag{2}$$

where ‘ ′ ’ denotes the matrix transpose, **X** is an $N \times K$ matrix of explanatory variables whose ($i$, $k$)-th element is given by $x_k(s_i)$, and **y** is an $N \times 1$ vector of the response variable whose $i$-th element is given by $y(s_i)$. $\mathbf{G}(s_i)$ is an $N \times N$ diagonal matrix whose $j$-th element is given by a geographically weighting function, $g(s_i, s_j)$. The specification of $g(s_i, s_j)$ is arbitrary, as long as the resulting covariance matrix is at least positive semi-definite. Wheeler and Calder (2007) and Wheeler and Waller (2009) applied the following exponential function form:

$$g(s_i, s_j) = \exp\left(-\frac{d(s_i, s_j)}{b}\right), \tag{3}$$

where $d(s_i, s_j)$ is the Euclidean distance between locations $s_i$ and $s_j$, and $b$ is the bandwidth parameter, which is large if coefficients have global scale spatial variation, and small if they have local scale spatial variation. A standard estimation procedure for the basic GWR specification is as follows: (1) the bandwidth is calculated based on cross-validation or a corrected AIC (see Fotheringham et al., 2002), and (2) $\boldsymbol{\beta}(s_i)$ is estimated by substituting the estimated bandwidth into Eqs. (2) and (3).

After Wheeler and Tiefelsdorf (2005) demonstrate that GWR coefficients essentially are collinear, active discussion shifted to regularized GWR models. For example, Wheeler (2007) proposes a ridge regularization-based GWR that replaces Eq. (2) with the following equation:

$$\hat{\boldsymbol{\beta}}(s_i) = (\mathbf{X}'\mathbf{G}(s_i)\mathbf{X} + \eta\mathbf{I}_K)^{-1}\mathbf{X}'\mathbf{G}(s_i)\mathbf{y}, \tag{4}$$

where $\eta$ is the ridge regularization parameter, and $\mathbf{I}_K$ is a $K \times K$ identity matrix. Wheeler (2009) and Gollini et al. (2015) extended the ridge GWR model to vary $\eta$ locally.

Specifically, they propose the locally compensated ridge GWR (LCR-GWR) model, whose estimator is as follows:

$$\hat{\boldsymbol{\beta}}(s_i) = (\mathbf{X}'\mathbf{G}(s_i)\mathbf{X} + \eta(s_i)\mathbf{I}_K)^{-1}\mathbf{X}'\mathbf{G}(s_i)\mathbf{y}, \quad (5)$$

where $\eta(s_i)$ is the ridge parameter for location $s_i$, and LCR-GWR calibrates $\eta(s_i)$ based on the degree of multicollinearity in the corresponding local model. Because $\eta(s_i)$ increases bias of the coefficient estimator, just like the standard ridge estimator, Gollini et al. (2015) suggest introducing $\eta(s_i)$ only for local models whose multicollinearity is excessive. The LCR-GWR model can be estimated as follows (Gollini et al., 2015): (1) the bandwidth and ridge parameters are estimated by cross-validation, and (2) $\boldsymbol{\beta}(s_i)$ is estimated by substituting them into Eq. (5).

## 3. ESF-based SVC models

Section 3.1 introduces the ESF approach 1, and section 3.2 presents an ESF-based SVC model, which we extend to a RE-ESF-based model in section 4.

*3.1. The ESF approach*

Moran ESF is based on the Moran coefficient (MC; see, Anselin and Rey, 1991), which is a spatial dependence diagnostic statistic formulated as follows[2]:

$$MC = \frac{N}{\mathbf{1}'\mathbf{C}\mathbf{1}} \frac{\mathbf{y}'\mathbf{MCMy}}{\mathbf{y}'\mathbf{My}}. \tag{6}$$

where $\mathbf{1}$ is an $N \times 1$ vector of ones, $\mathbf{y}$ is an $N \times 1$ vector of variable values, $\mathbf{C}$ is an $N \times N$ connectivity matrix whose diagonal elements are zero, and $\mathbf{M} = \mathbf{I}_N - \mathbf{11}'/N$ is an $N \times N$ matrix for double centering, where $\mathbf{I}_N$ is an $N \times N$ identity matrix. Note that $\mathbf{M}$ is replaced with $\mathbf{M}_X = \mathbf{I}_N - \mathbf{X}(\mathbf{X}'\mathbf{X})^{-1}\mathbf{X}'$ if $\mathbf{y}$ is a residual vector of a linear regression model. MC is positive if the sample values in $\mathbf{y}$ display positive spatial dependence, and negative if they display negative spatial dependence. The $l$-th eigenvector of $\mathbf{MCM}$, $\mathbf{e}_l$, describes the $l$-th map pattern explained by MC, while the set of eigenvectors of $\mathbf{MCM}$, $\mathbf{E}_{full} = \{\mathbf{e}_1, ..., \mathbf{e}_N\}$, provides all the possible distinct map pattern descriptions of latent spatial dependence, with each magnitude being indexed by its corresponding eigenvalue (Griffith, 2003).

[2] ESF also could be based on other indices, such as the Geary ratio (Geary, 1954).

ESF describes the latent map pattern in a georeferenced response variable $\mathbf{y}$, using a linear combination of eigenvectors, $\mathbf{E\gamma}$, where $\mathbf{E}$ is a matrix composed of $L$ eigenvectors in $\mathbf{E}_{full}$ ($L < N$), and $\boldsymbol{\gamma} = [\gamma_1, ..., \gamma_L]'$ is an $L \times 1$ coefficient vector. The linear ESF model is given by

$$\mathbf{y} = \mathbf{X\beta} + \mathbf{E\gamma} + \boldsymbol{\varepsilon}, \qquad \boldsymbol{\varepsilon} \sim N(\mathbf{0}_N, \sigma^2 \mathbf{I}_N), \tag{7}$$

where $\boldsymbol{\varepsilon}$ is an $N \times 1$ vector of disturbances, $\boldsymbol{\beta} = [\beta_1, ..., \beta_K]'$ is a $K \times 1$ parameter vector, $\mathbf{0}_N$ is an $N \times 1$ vector of zeros, and $\sigma^2$ is a variance parameter. Because Eq. (7) is in the form of the standard linear regression model, ordinary least squares (OLS) estimation is applicable for its parameter estimation[3]. The $L$ eigenvectors in $\mathbf{E}$ may be selected as follows (see, Chun et al., 2016): (a) eigenvectors corresponding to small eigenvalues, which explain the inconsequential level of spatial dependence, are removed, and (b) significant eigenvectors are chosen by applying a stepwise variable selection process to a predefined candidate set.

Many studies demonstrate the effectiveness of ESF in estimation and inference for $\boldsymbol{\beta}$ in the presence of spatial dependence (e.g., Chun, 2014; Griffith and Chun, 2014; Margaretic et al., 2015). For more details about ESF, see Griffith (2003), Griffith and Paelinck (2011), Griffith and Chun (2014), and Griffith and Chun (2016).

[3] Another approach includes the model selection procedure based on LASSO (Seya et al., 2015).

*3.2. ESF-based SVC models*

Griffith (2008) extended ESF to the following SVC model:

$$\mathbf{y} = \mathbf{X}\boldsymbol{\beta} + \sum_{k=1}^{K} \mathbf{x}_k \circ \mathbf{E}_k \boldsymbol{\gamma}_k + \boldsymbol{\varepsilon}, \qquad \boldsymbol{\varepsilon} \sim N(\mathbf{0}_N, \sigma^2 \mathbf{I}_N), \tag{8}$$

which also is expressed as

$$\mathbf{y} = \sum_{k=1}^{K} \mathbf{x}_k \circ \boldsymbol{\beta}_k^{ESF} + \boldsymbol{\varepsilon}, \qquad \boldsymbol{\varepsilon} \sim N(\mathbf{0}_N, \sigma^2 \mathbf{I}_N), \tag{9}$$

$$\boldsymbol{\beta}_k^{ESF} = \beta_k \mathbf{1} + \mathbf{E}_k \boldsymbol{\gamma}_k,$$

where $\mathbf{x}_k$ is an $N \times 1$ vector of the $k$-th explanatory variable (i.e., $k$-th column of $\mathbf{X}$), $\mathbf{E}_k$ is an $N \times L_k$ matrix composed of $L_k$ eigenvectors ($L_k < N$), $\boldsymbol{\gamma}_k$ is an $L_k \times 1$ coefficient vector, and '$\circ$' denotes the element-wise (Hadamard) product operator. $\boldsymbol{\beta}_k^{ESF} = \beta_k \mathbf{1} + \mathbf{E}_k \boldsymbol{\gamma}_k$ yields a vector of spatially varying coefficients, in which $\beta_k \mathbf{1}$ and $\mathbf{E}_k \boldsymbol{\gamma}_k$ represent the constant component and spatially varying component, respectively. The parameters can be estimated, as for the standard ESF specification, as follows: (a) eigenvectors corresponding to small eigenvalues are removed from each $\mathbf{E}_k$; (b) significant variables in $\mathbf{X}$, $\mathbf{x}_1 \circ \mathbf{E}_1$, ..., $\mathbf{x}_K \circ \mathbf{E}_K$ are selected by applying an OLS-based forward variable selection technique, and $\boldsymbol{\beta}$= $[\beta_1, ..., \beta_K]'$ and $\boldsymbol{\gamma}_k$ are then estimated; and, (c) $\hat{\boldsymbol{\beta}}_k^{ESF}$ = $\hat{\beta}_k \mathbf{1} + \mathbf{E}\hat{\boldsymbol{\gamma}}_k$ is calculated. Helbich and Griffith (2016) empirically demonstrated that spatial variation of the ESF-based coefficients can be significantly different from those for GWR.

## 4. RE-ESF-based SVC models

### *4.1. The RE-ESF approach*

While the conventional ESF model is a fixed effects model, Murakami and Griffith (2015) show the effectiveness of random effects-versions of ESF in terms of parameter inference. This section extends RE-ESF to a SVC model as follows:

$$\mathbf{y} = \mathbf{X}\boldsymbol{\beta} + \mathbf{E}\boldsymbol{\gamma} + \boldsymbol{\varepsilon}, \qquad \boldsymbol{\gamma} \sim N(\mathbf{0}_L, \sigma_\gamma^2 \boldsymbol{\Lambda}(\alpha)), \qquad \boldsymbol{\varepsilon} \sim N(\mathbf{0}_N, \sigma^2 \mathbf{I}_N), \tag{10}$$

where $\mathbf{0}_L$ is an $L \times 1$ vector of zeros, $\mathbf{E}$ is given by the subset of $L$ eigenvectors corresponding to positive eigenvalues (without applying the stepwise variable selection process), and $\mathbf{\Lambda}(\alpha)$ is an $L \times L$ diagonal matrix whose $l$-th element is $\lambda_l(\alpha) = (\Sigma_l\lambda_l/\Sigma_l\lambda_l^{\alpha})\lambda_l^{\alpha}$, where $\alpha$ and $\sigma_\gamma^2$ are parameters. A large $\alpha$ shrinks the coefficients of the non-principal eigenvectors strongly toward 0, and the resulting $\mathbf{E\gamma}$ describes a global map pattern. By contrast, when $\alpha$ is small, $\mathbf{E\gamma}$ describes a local map pattern. Thus, $\alpha$ controls the spatial smoothness of the underlying map pattern. RE-ESF has two interpretations (Murakami and Griffith, 2015): it describes a map pattern explained by MC, and it describes a Gaussian process after a rank reduction.

Eq. (10) can be rewritten as follows:

$$\mathbf{y} = \mathbf{X\beta} + \mathbf{EV}(\mathbf{\theta})\mathbf{u} + \mathbf{\varepsilon}, \qquad \mathbf{u} \sim N(\mathbf{0}_L, \sigma^2\mathbf{I}_L), \qquad \mathbf{\varepsilon} \sim N(\mathbf{0}_N, \sigma^2\mathbf{I}_N), \qquad (11)$$

where $\mathbf{\theta} = \{\sigma_\gamma^2/\sigma^2, \alpha\}$, $\mathbf{I}_L$ is an $L \times L$ identity matrix, and $\mathbf{V}(\mathbf{\theta})$ is a diagonal matrix whose $l$-th element is $(\sigma_\gamma/\sigma)\lambda_l(\alpha)^{1/2}$. Note that $\mathbf{V}(\mathbf{\theta})\mathbf{u}$ in Eq. (11) equals $\mathbf{\gamma}$.

The parameters in Eq. (11) (or Eq. (10)) are estimated by using the residual maximum likelihood (REML) method of Bates (2010). Following his specification, the likelihood function is defined by $loglik(\mathbf{\beta}, \mathbf{\theta}) = \int p(\mathbf{y}, \mathbf{u} \mid \mathbf{\beta}, \mathbf{\theta}) d\mathbf{u}$, and the restricted log-likelihood by $loglik_R(\boldsymbol{\theta}) = \int loglik(\boldsymbol{\beta}, \boldsymbol{\theta}) d\boldsymbol{\beta}$.

The estimation procedure is summarized as follows: (a) $\boldsymbol{\theta}$ is estimated by maximizing the restricted log-likelihood Eq. (12) with the plugins of Eqs. (13) and (14); (b) $\boldsymbol{\beta}$ and $\boldsymbol{\gamma} = \mathbf{V}(\boldsymbol{\theta})\mathbf{u}$ are estimated by substituting the estimated $\boldsymbol{\theta}$ into Eq. (14); and, (c) $\sigma^2$ is estimated by substituting the estimated parameters into Eq. (15). In other words,

$$loglik_R(\boldsymbol{\theta}) = -\frac{1}{2}\log\left\|\begin{bmatrix} \mathbf{X}'\mathbf{X} & \mathbf{X}'\mathbf{E}\mathbf{V}(\boldsymbol{\theta}) \\ \mathbf{V}(\boldsymbol{\theta})\mathbf{E}'\mathbf{X} & \mathbf{V}(\boldsymbol{\theta})^2 + \mathbf{I}_L \end{bmatrix}\right\| - \frac{N-K}{2}\left(1 + \log\left(\frac{2\pi d(\boldsymbol{\theta})}{N-K}\right)\right), \quad (12)$$

$$d(\boldsymbol{\theta}) = \min_{\boldsymbol{\beta},\mathbf{u}} \ \| \mathbf{y} - \mathbf{X}\boldsymbol{\beta} - \mathbf{E}\mathbf{V}(\boldsymbol{\theta})\mathbf{u} \|^2 + \| \mathbf{u} \|^2, \quad (13)$$

$$\begin{bmatrix} \hat{\boldsymbol{\beta}} \\ \hat{\mathbf{u}} \end{bmatrix} = \begin{bmatrix} \mathbf{X}'\mathbf{X} & \mathbf{X}'\mathbf{E}\mathbf{V}(\boldsymbol{\theta}) \\ \mathbf{V}(\boldsymbol{\theta})\mathbf{E}'\mathbf{X} & \mathbf{V}(\boldsymbol{\theta})^2 + \mathbf{I}_L \end{bmatrix}^{-1} \begin{bmatrix} \mathbf{X}'\mathbf{y} \\ \mathbf{V}(\boldsymbol{\theta})\mathbf{E}'\mathbf{y} \end{bmatrix}, \text{ and} \quad (14)$$

$$\hat{\sigma}^2 = \frac{\| \mathbf{y} - \mathbf{X}\boldsymbol{\beta} - \mathbf{E}\mathbf{V}(\boldsymbol{\theta})\mathbf{u} \|^2}{N-K}, \quad (15)$$

where $\| \bullet \|^2$ denotes the $L_2$-norm of a vector $\bullet$, and $\mathbf{V}(\boldsymbol{\theta})^2 = \mathbf{V}(\boldsymbol{\theta})\mathbf{V}(\boldsymbol{\theta}) = \mathbf{V}(\boldsymbol{\theta})\mathbf{E}'\mathbf{E}\mathbf{V}(\boldsymbol{\theta})$. The computational complexity of Eq. (12) is $O((K+L)^3)$, which is smaller than the complexity of the likelihood maximization in standard spatial statistical models ($O(N^3)$). Murakami and Griffith (2015) reveal that RE-ESF outperforms ESF in terms of not only the effectiveness of estimating and inferring $\boldsymbol{\beta}$, but also computation time.

Although similar models have been used in the statistics literature (e.g., Hughes and Haran, 2013; Johnson et al., 2013; Lee and Barran, 2015), they use **E** generated not from **MCM**, but from $\mathbf{M}_X\mathbf{C}\mathbf{M}_X$. This is because the latter specification eliminates confounders between **X** and **E**, and stabilizes the parameter estimates. However, the latter leads to a serious underestimation of the standard errors of $\boldsymbol{\beta}$ (Murakami and Griffith, 2015), and introduces a Type I error if factors omitted from the model are correlated with **X**, as is known in spatial econometrics (LeSage and Pace, 2009; Griffith and Chun, 2016). Hence, we utilize the former specification.

### *4.2. RE-ESF-based SVC models*

We combine the RE-ESF model and the ESF-based SVC model (Eq. (9)):

$$\mathbf{y} = \sum_{k=1}^{K} \mathbf{x}_k \circ \boldsymbol{\beta}_k^{R-ESF} + \boldsymbol{\varepsilon}, \qquad \boldsymbol{\varepsilon} \sim N(\mathbf{0}_N, \sigma^2 \mathbf{I}_N), \tag{16}$$

$$\boldsymbol{\beta}_k^{R-ESF} = \beta_k \mathbf{1} + \mathbf{E}\boldsymbol{\gamma}_k, \qquad \boldsymbol{\gamma}_k \sim N(\mathbf{0}_L, \sigma_{k(\gamma)}^2 \boldsymbol{\Lambda}(\alpha_k)),$$

where $\alpha_k$ is a parameter that controls the spatial smoothness of the $k$-th coefficients, and $\sigma^2_{k(\gamma)}$ controls the variance. The $k$-th coefficients consist of the fixed constant, $\beta_k\mathbf{1}$, and random spatially varying components, $\mathbf{E}\boldsymbol{\gamma}_k$.

Eq. (16) can be expressed as (see Eqs. (8) and (9))

$$\mathbf{y} = \mathbf{X}\boldsymbol{\beta} + \tilde{\mathbf{E}}\tilde{\boldsymbol{\gamma}} + \boldsymbol{\varepsilon}, \qquad \boldsymbol{\varepsilon} \sim N(\mathbf{0}_N, \sigma^2\mathbf{I}_N), \tag{17}$$

where

$$\tilde{\mathbf{E}} = \begin{bmatrix} \mathbf{x}_1 \circ \mathbf{E} & \cdots & \mathbf{x}_K \circ \mathbf{E} \end{bmatrix}, \quad \tilde{\boldsymbol{\gamma}} = \begin{bmatrix} \boldsymbol{\gamma}_1 \\ \vdots \\ \boldsymbol{\gamma}_K \end{bmatrix} \sim N\left[ \begin{pmatrix} \mathbf{0}_L \\ \vdots \\ \mathbf{0}_L \end{pmatrix}, \begin{pmatrix} \sigma_{1(\gamma)}^2 \boldsymbol{\Lambda}(\alpha_1) & & \\ & \ddots & \\ & & \sigma_{K(\gamma)}^2 \boldsymbol{\Lambda}(\alpha_K) \end{pmatrix} \right].$$

Eq. (17) essentially is identical to Eq. (10). Hence, it is further expanded similar to the expansion of Eq. (10) to Eq. (11):

$$\mathbf{y} = \mathbf{X}\boldsymbol{\beta} + \tilde{\mathbf{E}}\tilde{\mathbf{V}}(\boldsymbol{\Theta})\tilde{\mathbf{u}} + \boldsymbol{\varepsilon}, \qquad \tilde{\mathbf{u}} \sim N(\mathbf{0}_{LK}, \sigma^2\mathbf{I}_{LK}), \qquad \boldsymbol{\varepsilon} \sim N(\mathbf{0}_N, \sigma^2\mathbf{I}_N), \tag{18}$$

$$\tilde{\mathbf{V}}(\boldsymbol{\Theta}) = \begin{bmatrix} \mathbf{V}(\boldsymbol{\theta}_1) & & \\ & \ddots & \\ & & \mathbf{V}(\boldsymbol{\theta}_K) \end{bmatrix}, \qquad \tilde{\mathbf{u}} = \begin{bmatrix} \mathbf{u}_1 \\ \vdots \\ \mathbf{u}_K \end{bmatrix},$$

where $\boldsymbol{\Theta} = \{\boldsymbol{\theta}_1, ..., \boldsymbol{\theta}_K\}$, $\boldsymbol{\theta}_k = \{\sigma_{k(\gamma)}{}^2/\sigma^2, \alpha_k\}$, $\mathbf{0}_{LN}$ is an $L_K \times 1$ vector of zeros, $\mathbf{I}_{LN}$ is an $L_K \times L_K$ identity matrix, and $\mathbf{V}(\boldsymbol{\theta}_k)$ is a diagonal matrix whose $l$-th element is $(\sigma_{\gamma(\gamma)}/\sigma)\lambda_l(\alpha_k)^{1/2}$.

Because Eq. (17) is identical to the RE-ESF model, Eq. (10), the REML estimation for RE-ESF is readily applicable to the proposed model. The estimation procedure is summarized as follows: (a) $\boldsymbol{\Theta}$ is estimated by maximizing the profile restricted log-likelihood, Eq. (19), with the plugins of Eqs. (20) and (21); (b) $\boldsymbol{\beta}$ and $\tilde{\boldsymbol{\gamma}} = \tilde{\mathbf{V}}(\boldsymbol{\Theta})\tilde{\mathbf{u}}$ are estimated by substituting the estimated $\boldsymbol{\Theta}$ into Eq. (21); and, (c) $\sigma^2$ is estimated by substituting the estimated parameters into Eq. (22). In other words,

$$loglik_R(\boldsymbol{\Theta}) = -\frac{1}{2}\log\left|\begin{bmatrix} \mathbf{X}'\mathbf{X} & \mathbf{X}'\tilde{\mathbf{E}}\tilde{\mathbf{V}}(\boldsymbol{\Theta}) \\ \tilde{\mathbf{V}}(\boldsymbol{\Theta})\tilde{\mathbf{E}}'\mathbf{X} & \tilde{\mathbf{V}}(\boldsymbol{\Theta})\tilde{\mathbf{E}}'\tilde{\mathbf{E}}\tilde{\mathbf{V}}(\boldsymbol{\Theta}) + \mathbf{I}_{LK} \end{bmatrix}\right| - \frac{N-K}{2}\left(1 + \log\left(\frac{2\pi\tilde{d}(\boldsymbol{\Theta})}{N-K}\right)\right), \quad (19)$$

$$\tilde{d}(\boldsymbol{\Theta}) = \min_{\boldsymbol{\beta},\tilde{\mathbf{u}}} \|\mathbf{y} - \mathbf{X}\boldsymbol{\beta} - \tilde{\mathbf{E}}\tilde{\mathbf{V}}(\boldsymbol{\Theta})\tilde{\mathbf{u}}\|^2 + \|\tilde{\mathbf{u}}\|^2, \quad (20)$$

$$\begin{bmatrix} \hat{\boldsymbol{\beta}} \\ \hat{\tilde{\mathbf{u}}} \end{bmatrix} = \begin{bmatrix} \mathbf{X}'\mathbf{X} & \mathbf{X}'\tilde{\mathbf{E}}\tilde{\mathbf{V}}(\boldsymbol{\Theta}) \\ \tilde{\mathbf{V}}(\boldsymbol{\Theta})\tilde{\mathbf{E}}'\mathbf{X} & \tilde{\mathbf{V}}(\boldsymbol{\Theta})\tilde{\mathbf{E}}'\tilde{\mathbf{E}}\tilde{\mathbf{V}}(\boldsymbol{\Theta}) + \mathbf{I}_{LK} \end{bmatrix}^{-1} \begin{bmatrix} \mathbf{X}'\mathbf{y} \\ \tilde{\mathbf{V}}(\boldsymbol{\Theta})\tilde{\mathbf{E}}'\mathbf{y} \end{bmatrix}, \text{ and} \quad (21)$$

$$\hat{\sigma}^2 = \frac{\|\mathbf{y} - \mathbf{X}\boldsymbol{\beta} - \tilde{\mathbf{E}}\tilde{\mathbf{V}}(\boldsymbol{\Theta})\tilde{\mathbf{u}}\|^2}{N-K}. \quad (22)$$

Although the REML estimation requires a determinant calculation, computational complexity is only $O((K+KL)^3)$, which can be decreased by reducing the number of eigenvectors in $\mathbf{E}$. The computational burden also can be reduced by replacing some $\boldsymbol{\beta}_k^{R\text{-}ESF} = \beta_k\mathbf{1} + \mathbf{E}\boldsymbol{\gamma}_k$ with $\beta_k\mathbf{1}$, which means restricting some coefficients to be constants across a given geographic landscape.

The variance-covariance matrices of the coefficients are

$$Cov\begin{bmatrix}\boldsymbol{\beta}\\ \tilde{\boldsymbol{\gamma}}\end{bmatrix}=Cov\begin{bmatrix}\boldsymbol{\beta}\\ \tilde{\mathbf{V}}(\boldsymbol{\Theta})\tilde{\mathbf{u}}_k\end{bmatrix}=\sigma^2\begin{bmatrix}\mathbf{X}'\mathbf{X} & \mathbf{X}'\tilde{\mathbf{E}}\\ \tilde{\mathbf{E}}'\mathbf{X} & \tilde{\mathbf{E}}'\tilde{\mathbf{E}}+\tilde{\mathbf{V}}(\boldsymbol{\Theta})^{-2}\end{bmatrix}^{-1}, \tag{23}$$

where $\tilde{\mathbf{V}}(\boldsymbol{\Theta})^{-2}$ is the inverse of $\tilde{\mathbf{V}}(\boldsymbol{\Theta})^{2}$. Because $\tilde{\mathbf{V}}(\boldsymbol{\Theta})^{2}$ is a diagonal matrix, its inverse is easily calculated. As for $\boldsymbol{\beta}_k^{R-ESF} = \beta_k\mathbf{1} + \mathbf{E}\boldsymbol{\gamma}_k$, the variance of the constant component, $\beta_k\mathbf{1}$, is estimated in Eq. (23). The covariance matrix of the spatially varying components, $\mathbf{E}\boldsymbol{\gamma}_k$, is estimated as follows:

$$Cov[\mathbf{E}\boldsymbol{\gamma}_k]=\mathbf{E}\boldsymbol{\gamma}_k\boldsymbol{\gamma}_k'\mathbf{E}'=\mathbf{E}Cov[\boldsymbol{\gamma}_k]\mathbf{E}', \tag{24}$$

where $Cov[\boldsymbol{\gamma}_k]$, which is the covariance matrix of $\boldsymbol{\gamma}_k$, is a sub-matrix of $Cov[\tilde{\boldsymbol{\gamma}}]$ in Eq.(23). The diagonals of Eq. (24) are useful to test if $\boldsymbol{\beta}_k^{R-ESF} = \beta_k\mathbf{1} + \mathbf{E}\boldsymbol{\gamma}_k$ has statistically significant spatial variation.

A problem is how to estimate $\boldsymbol{\Theta}$ efficiently. For example, when five explanatory variables are considered, we need to optimize 10 parameters in $\{\sigma_{1}^{2}{}_{(\gamma)}, ..., \sigma_{5}^{2}{}_{(\gamma)}, \alpha_1, ..., \alpha_5\}$ simultaneously, which can be computationally expensive. Hence, in addition to simultaneous estimation, we apply an approximation that estimates the coefficient's variance parameters, $\sigma_{k}^{2}{}_{(\gamma)}$s, first, and the spatial smoothness parameters, $\alpha_k$s, thereafter. In the first step, we impose $\alpha_k = 1$, which implicitly has been assumed in RE-ESF-type models (e.g., Hughes and Haran, 2013). Assuming a unique value for each $\alpha_k$, which implies the same degree of spatial smoothness for each coefficient, is another way to increase computational efficiency. Section 6 compares the effectiveness of these simplifications.

## 5. Properties of RE-ESF-based SVC model

This section clarifies advantages and disadvantages of our SVC model by comparing it with the ESF-based SVC model (section 5.1), GWR models (section 5.2), and the B-SVC model of Gelfand (2003) (section 5.3).

### *5.1. A comparison with the ESF-based model*

Both the ESF-based model and our model describe their *k*-th coefficients using $\beta_k \mathbf{1} + \mathbf{E}\boldsymbol{\gamma}_k$. The ESF approach regards $\mathbf{E}\boldsymbol{\gamma}_k$ as fixed effects, whereas ours considers it as random effects, where $\boldsymbol{\gamma}_k \sim N(\mathbf{0}_L, \sigma_{k(\gamma)}^2\boldsymbol{\Lambda}(\alpha_k))$. Our specification has additional variance parameters, $\sigma_{k(\gamma)}^2$ and $\alpha_k$. They shrink $\mathbf{E}\boldsymbol{\gamma}_k$ strongly toward zero when $\sigma_{k(\gamma)}^2$ is small and $\alpha_k$ is large. Owing to these parameters, our estimator might be more robust to multicollinearity than the estimator of ESF, which is a fundamental problem in SVC models (Wheeler and Tiefelsdorf, 2005).

The parameter $\alpha_k$ also controls the spatial smoothness of each varying coefficient. A large $\alpha_k$ shrinks the coefficients $\gamma_{k,l}$ corresponding to the non-principal eigenvectors strongly toward zero, where $\gamma_{k,l}$ is the $l$-th element of $\boldsymbol{\gamma}_k$. As a result, $\mathbf{E}\boldsymbol{\gamma}_k$ has a global (smoother) map pattern. Interestingly, $\alpha_k$ is interpretable in terms of MC. $MC[\mathbf{E}\boldsymbol{\gamma}_k]$ can be calculated by substituting $\mathbf{E}\boldsymbol{\gamma}_k$ into Eq. (6) as follows (see Griffith, 2003):

$$MC[\mathbf{E}\boldsymbol{\gamma}_k] = \frac{N}{\mathbf{1}'\mathbf{C}\mathbf{1}} \frac{1}{\sum_{l=1}^{L} \gamma_{k,l}^2} \sum_{l=1}^{L} \gamma_{k,l}^2 \lambda_l \ . \tag{25}$$

$MC[\mathbf{E}\boldsymbol{\gamma}_k]$ grows as $\gamma_{k,l}$ corresponding to large (in absolute value) eigenvalues increases compared with those corresponding to medium-small (in absolute value) eigenvalues. This means that $MC[\mathbf{E}\boldsymbol{\gamma}_k]$ increases if $\alpha_k$, which shrinks $\gamma_{k,l}$ corresponding to medium-small (absolute) eigenvalues, is large. In particular, $MC[\mathbf{E}\boldsymbol{\gamma}_k]$ takes its maximum value if $\alpha_k = \infty$ because $\gamma_{k,2} = \gamma_{k,3} ... = \gamma_{k,L} = 0$. By contrast, if $\alpha_k = 0$, $\sigma_{k(\gamma)}^2$ shrinks all coefficients equally. In this case, the estimator of $\gamma_{k,l}$ yields the standard ridge regression estimator. In short, $\alpha_k$ is an MC-based shrinkage parameter that intensifies the underlying spatial dependence of $\boldsymbol{\beta}_k^{R-ESF} = \beta_k\mathbf{1} + \mathbf{E}\boldsymbol{\gamma}_k$.

Computational efficiency is another advantage of our approach. While we apply REML estimation, the ESF-based SVC model applies stepwise variable selection, which is computationally costly especially for massively large datasets. An advantage of ESF is its simplicity (Griffith, 2003). If the efficiency of ESF is compatible with RE-ESF, ESF remains a useful alternative.

*5.2. A comparison with GWR models*

A major advantage of our model relative to GWR is its capability of allowing different spatial smoothness of SVCs. GWR studies usually assume the same degree of spatial smoothness for each coefficient, which is unlikely in many real-world situations. Moreover, our approach estimates coefficients based on a global estimation, whereas GWR iterates with local estimations. The global estimation that considers all observations might be more robust than local estimations that consider nearby observations only. Indeed, the efficiency of local estimations depends on the rank sufficiency and collinearity of the (geographically weighted) explanatory variables around each site. Our global estimation is not compromised by such problems.

By contrast, GWR is simpler and easier to extend for non-Gaussian data modeling, spatial interpolation, and other purposes (Fotheringham et al., 2002; Nakaya et al., 2005). Besides, GWR is applicable to a large data set, and can be made faster with parallel computing (Harris et al., 2010), whereas our model is not parallelizable because it requires an eigen-decomposition. Furthermore, GWR models are easily implemented (e.g., using the Spatial Statistics Toolbox in ArcGIS (http://www.esri.com/), or spgwr, gwrr, and GWmodel in the R packages (see Gollini et al., 2015)). Our model needs to be extended to overcome these disadvantages.

### *5.3. A comparison with the B-SVC models*

The B-SVC model is formulated as follows:

$$\mathbf{y} = \beta_1 \mathbf{1} + \sum_{k=2}^{K} \mathbf{x}_k \circ \boldsymbol{\beta}_k^{B-SVC} + \mathbf{e}_1, \qquad \mathbf{e}_1 \sim N(\mathbf{0}_N, \delta_1^2 \mathbf{C}(\boldsymbol{\theta}_1) + \tau_1^2 \mathbf{I}_N), \tag{26}$$

$$\boldsymbol{\beta}_k^{B-SVC} = \beta_k \mathbf{1} + \mathbf{M}\mathbf{e}_k, \qquad \mathbf{e}_k \sim N(\mathbf{0}_N, \delta_k^2 \mathbf{C}(\boldsymbol{\theta}_k) + \tau_k^2 \mathbf{I}_N), \tag{27}$$

where $\delta^2{}_k$ and $\tau^2{}_k$ are variance parameters, and $\boldsymbol{\theta}_k$ is a collection of parameters characterizing the covariance matrix $\mathbf{C}(\boldsymbol{\theta}_k)$. B-SVC describes both SVCs with [a constant term: $\beta_k\mathbf{1}$] + [a centered Gaussian process, $\mathbf{M}\mathbf{e}_k$], and residuals with another Gaussian process.

Murakami and Griffith (2015) show that $\mathbf{M}\mathbf{e}_k \sim N(\mathbf{0}_N, \delta_k^2\mathbf{M}\mathbf{C}(\boldsymbol{\theta}_k)\mathbf{M}+\tau^2{}_k\mathbf{M})$ can be expanded as follows, after a rank reduction[4]:

$$\mathbf{M}\mathbf{e}_k = \mathbf{E}\boldsymbol{\gamma}_k + \boldsymbol{\varepsilon}_k, \qquad \boldsymbol{\gamma}_k \sim N(\mathbf{0}_L, \delta_k^2 \boldsymbol{\Lambda}), \qquad \boldsymbol{\varepsilon}_k \sim N(\mathbf{0}_N, \tau_k^2 \mathbf{I}_N), \tag{28}$$

where $\boldsymbol{\theta}_k$ is omitted for simplicity. Eqs. (27) and (28) indicate that $\boldsymbol{\beta}_k^{B-SVC} = \beta_k\mathbf{1} + \mathbf{E}\boldsymbol{\gamma}_k + \boldsymbol{\varepsilon}_k$ (after a rank reduction), whereas our model yields $\boldsymbol{\beta}_k^{R-ESF} = \beta_k\mathbf{1} + \mathbf{E}\boldsymbol{\gamma}_k$ (see Eq. (16)). $\boldsymbol{\beta}_k^{R-ESF}$, which does not include $\boldsymbol{\varepsilon}_k$, captures a smoother map pattern than $\boldsymbol{\beta}_k^{B-SVC}$. The difference between $\boldsymbol{\beta}_k^{R-ESF}$ and $\boldsymbol{\beta}_k^{B-SVC}$ arises because our model is based on the MC, which does not consider variances within each sample, whereas the B-SVC model describes Gaussian processes, which capture within sample variance with $\delta_k^2$ and $\sigma_k^2$.

[4] Here, $\mathbf{M}\mathbf{M}' = \mathbf{M}$ is used. It holds because $\mathbf{M}$ is a symmetric and idempotent matrix.

Suppose that $\mathbf{x}_1$ is a constant; then, our model, Eq. (16), can be expanded using Eqs. (27) and (28):

$$\mathbf{y} = \boldsymbol{\beta}_1^{R-ESF} + \sum_{k=2}^{K} \mathbf{x}_k \circ \boldsymbol{\beta}_k^{R-ESF} + \boldsymbol{\varepsilon}, \qquad \boldsymbol{\varepsilon} \sim N(\mathbf{0}_N, \sigma^2 \mathbf{I}_N).$$

$$\mathbf{y} = \beta_1 \mathbf{1} + \sum_{k=2}^{K} \mathbf{x}_k \circ \boldsymbol{\beta}_k^{R-ESF} + \mathbf{E}\boldsymbol{\gamma}_1 + \boldsymbol{\varepsilon}, \qquad \boldsymbol{\varepsilon} \sim N(\mathbf{0}_N, \sigma^2 \mathbf{I}_N).$$

$$= \beta_1 \mathbf{1} + \sum_{k=2}^{K} \mathbf{x}_k \circ \boldsymbol{\beta}_k^{R-ESF} + \mathbf{M}\mathbf{e}_1, \qquad \mathbf{e}_1 \sim N(\mathbf{0}_N, \delta_1^2 \mathbf{C} + \sigma^2 \mathbf{I}_N), \qquad (29)$$

Thus, our model is a variant of the B-SVC model whose $\boldsymbol{\beta}_k^{B-SVC}$ is replaced with $\boldsymbol{\beta}_k^{R-ESF}$, and the Gaussian process, $\mathbf{e}_1$, with a centered Gaussian process, $\mathbf{M}\mathbf{e}_1$.

A difference between these models is that while our model is estimated by the REML method, the B-SVC model must be estimated with MCMC. Because MCMC is robust, even if a sample size is small, the B-SVC model is preferable for small-to-medium size samples. However, MCMC is computationally expensive, particularly when different degrees of spatial smoothness are allowed for each coefficient (Finley, 2011). Therefore, our model is more suitable for medium-to-large size samples. Because our method does not require iterative sampling, unlike MCMC, it is preferable to B-SVC in terms of simplicity, too.

## 6. Results from a Monte Carlo simulation experiment

This section summarizes a Monte Carlo simulation experiment comparing our model with GWR models and the ESF-based model in terms of SVCs estimation accuracy and computational efficiency.

### *6.1. Outline*

This section compares the conventional GWR, LCR-GWR and ESF-based SVC models with our RE-ESF-based model. We also compare the following approximations of RE-ESF: the RE-ESF that estimates $\sigma^2_{k(\gamma)}$s first and $\alpha_k$s thereafter (RE-ESF (A1)), and the RE-ESF whose $\alpha_k$s are assumed to be uniform (RE-ESF (A2)).

The exponential model, Eq. (3), is used to evaluate the geographical weights in the GWR and LCR-GWR models. Regarding RE-ESF, a similar exponential model, Eq.(30), is used to evaluate the $(i, j)$-th element of the proximity matrix **C**, $c_i$:

$$c_{i,j} = \begin{cases} \exp(-d_{i,j} / r) & if \;\; i \neq j \\ 0 & otherwise. \end{cases} \quad (30)$$

Following Dray et al. (2006), the range parameter $r$ is given by the maximum distance in the minimum spanning tree connecting all sample sites. **E** in RE-ESF consists of the eigenvectors corresponding to positive eigenvalues. The same eigenvectors are regarded as candidates to be entered into the ESF model, and they are selected by the adjusted-$R^2$ based forward variable selection technique. This distance-based ESF often is called Moran's eigenvector maps, a popular approach in ecology (see, Dray et al., 2006; Griffith and Peres-Neto, 2006; Legendre and Legendre, 2012). Regarding ESF, to cope with multicollinearity, variables with variance inflation factors (VIFs) above 10 are excluded from the candidates in each variable selection step. As for LCR-GWR, following Gollini et al. (2015), the ridge term is introduced only for local models whose condition number exceeds 30.

Initially, we used ESF and RE-ESF to generate synthetic response variables. However, because results were too optimistic for our model, a situation not likely to occur in real world cases (in many cases, more than half of the estimation error is reduced compared with GWR), we do not report these result. Instead, we generate data

using Eq.(31):

$$\mathbf{y} = \boldsymbol{\beta}_0 + \mathbf{x}_1 \circ \boldsymbol{\beta}_1 + \mathbf{x}_2 \circ \boldsymbol{\beta}_2 + \boldsymbol{\varepsilon}, \qquad \boldsymbol{\varepsilon} \sim N(\mathbf{0}_N, 2^2\mathbf{I}_N) \tag{31}$$

$$\boldsymbol{\beta}_0 = \mathbf{1} + \mathbf{M}\widetilde{\mathbf{C}}(1)\boldsymbol{\varepsilon}_0 \qquad \boldsymbol{\beta}_1 = (-2)\mathbf{1} + \mathbf{M}\widetilde{\mathbf{C}}(r_1)\boldsymbol{\varepsilon}_1 \qquad \boldsymbol{\beta}_2 = (0.5)\mathbf{1} + \mathbf{M}\widetilde{\mathbf{C}}(r_2)\boldsymbol{\varepsilon}_2$$

where $\boldsymbol{\varepsilon}_k \sim N(\mathbf{0}_N, \mathbf{I}_N)$, and $\widetilde{\mathbf{C}}(r_k)$ is a matrix that row-standardizes $\mathbf{I}_N + \mathbf{C}(r_k)$, where $\mathbf{C}(r_k)$ is $\mathbf{C}$ whose range parameter is $r_k$. Eq. (31) models $\boldsymbol{\beta}_0$, $\boldsymbol{\beta}_1$ and $\boldsymbol{\beta}_2$ using centered moving average processes whose spatial smoothness is controlled by the ranges $r_0 = 1$, $r_1$ and $r_2$, respectively. $\mathbf{x}_1$, whose coefficients take –2 on average, accounts for more of the variation in $\mathbf{y}$, whereas $\mathbf{x}_2$, whose coefficients take 0.5 on average, accounts for less variation. Because GWR and (RE-)ESF coefficients are based on spatial smoothing with $c_{i,j} = \exp(-d_{i,j}/r)$, both types of coefficients describe $\boldsymbol{\beta}_k$ in Eq.(31) reasonably well.

The covariates in Eq. (31) are generated from Eq. (32):

$$\mathbf{x}_k = (1 - w_s)\mathbf{M}\mathbf{e}_{k(ns)} + w_s\mathbf{M}\widetilde{\mathbf{C}}(1)\mathbf{e}_{k(s)}, \quad \mathbf{e}_{k(ns)} = N(\mathbf{0}_N, \mathbf{I}_N), \quad \mathbf{e}_{k(s)} = N(\mathbf{0}_N, \mathbf{I}_N). \tag{32}$$

Eq. (32) assumes that $\mathbf{x}_k$ equals [the centered disturbance, $\mathbf{M}\mathbf{e}_{k(ns)}$] + [the centered spatially dependent component, $\mathbf{M}\widetilde{\mathbf{C}}(1)\mathbf{e}_{k(s)}$], whose contribution ratios are $1 - w_s$ and $w_s$, respectively. $\mathbf{x}_k$ has strong spatial dependence when $w_s$ is near 1. Some studies (e.g.,

Hughes and Haran, 2013) reveal that coefficient estimates tend to be unstable when explanatory variables are spatially dependent. This is because spatially dependent explanatory variables can confound with spatially dependent errors. However, no study has examined the extent to which such spatial confounding influences the coefficient estimates. We examine it by varying the intensity of spatial dependence in $\mathbf{x}_k$ with $w_s$.

Because we do not know how to control the degrees of spatial confounding and multicollinearity simultaneously, this simulation focuses on only the former, whose discussion is limited.

The response variable and covariates are generated on $N$ sample sites whose two geocoded coordinates are given by two random samples from $N(\mathbf{0}_N, \mathbf{I}_N)$[5]. Then, SVC models are fitted to these variables, and $\boldsymbol{\beta}_0$, $\boldsymbol{\beta}_1$ and $\boldsymbol{\beta}_2$ are estimated iteratively while varying the sample size $N \in \{50, 150, 400\}$, the ratio of the spatial dependence component in $\mathbf{x}_k$, $w_s \in \{0.0, 0.4, 0.8\}$, and the spatial smoothness of the coefficients: $\boldsymbol{\beta}_2$ and $\boldsymbol{\beta}_3$; $(r_2, r_3)$ ={(0.5, 1.0), (1.0, 0.5), (1.0, 1.0), (1.0, 2.0), (2.0, 1.0)}. In each case,

5 An assumption of $N(\mathbf{0}_N, \mathbf{I}_N)$ implies fewer samples near periphery areas. It is likely for many socioeconomic data including land price data, which typically have fewer samples in suburban areas.

estimations are iterated 200 times. These simulations are performed using R version 3.1.1 (https://cran.r-project.org/) on a 64 bit PC whose memory is 48 GB.

### *6.2. Results*

The estimation accuracy is evaluated by the root mean squared error (RMSE),

$$RMSE[\hat{\beta}_{k,i}] = \sqrt{\frac{1}{N}\sum_{i=1}^{N}(\hat{\beta}_{k,i} - \beta_{k,i})^2}\ , \tag{33}$$

where $\beta_{k,i}$ is the $i$-th element of the true $\boldsymbol{\beta}_k$, and $\hat{\beta}_{k,i}$ is the estimate. Tables 1–3 summarize the RMSEs in cases of $N$ = 50, 150 and 400, respectively.

**[Table 1 around here]**

**[Table 2 around here]**

**[Table 3 around here]**

The estimates of RE-ESF are more accurate than those of GWR and LCR-GWR for a medium-to-large sample size ($N$ = 150 or 400). This tendency is significant if the explanatory variables are spatially dependent (i.e., $w_s$ is large). By contrast, when $N = 50$, although the RE-ESF is still better than GWR models, their gaps are relatively small because RE-ESF relies on an REML estimation, which is less efficient for small samples.

Among RE-ESF models, the RE-ESF without an approximation is the best in many cases. By contrast, the RMSEs of RE-ESF (A1) are very close to those of RE-ESF, and the tendency is prominent when $N = 400$. Thus, RE-ESF (A1) would be a good alternative as long as the sample size is not small.

$\boldsymbol{\beta}_2$ conveys relatively minor effects. $RMSE[\hat{\boldsymbol{\beta}}_2]$ tends to be small in RE-ESF (A2), which assumes constant $\alpha_k$s, rather than RE-ESF and RE-ESF (A1), which assume non-constant $\alpha_k$s. In other words, the estimation of the coefficient smoothness parameters ($\alpha_k$s) can fail to capture the spatial variation of the SVCs, accounting for a small portion of variations in **y**. Nevertheless, the gaps in their RMSEs are marginal, and their RMSEs are smaller than those of the GWR and LCR-GWR models.

$\boldsymbol{\beta}_1$ describes relatively strong impacts. The $RMSE[\hat{\boldsymbol{\beta}}_1]$ of RE-ESF and RE-ESF

(A1) are smaller than those of RE-ESF (A2). This tendency is substantial when the covariates have strong spatial dependence (i.e., $w_s$ is large). This result suggests that the coefficient smoothness parameters in RE-ESF and RE-ESF (A1) play an important role in appropriately capturing SVCs, accounting for a large portion of variations in $\mathbf{y}$.

These tables also show the large RMSEs of the ESF coefficients. The result is possibly because ESF does not consider eigenvalues, which act as deflators for coefficients for eigenvectors corresponding to small (in absolute value) eigenvalues in our model.

In each model, RMSE increases in the presence of strong spatial dependence in the covariates, which can confound with spatially dependence in residuals. This result reveals the importance of considering the confounding factor typically ignored in SVC-related studies. Increases in the RMSEs are relatively small in RE-ESF and RE-ESF (A1), including the coefficient smoothness parameter, $\alpha_k$, which thus might be helpful in mitigating this problem.

Table 4 summarizes mean bias, which is defined as follows:

$$mean\, bias = \frac{1}{N}\sum_{i=1}^{N}(\hat{\beta}_{k,i} - \beta_{k,i}). \qquad (34)$$

Eq. (34) is positive if $\hat{\boldsymbol{\beta}}_k$ is positively biased, and negative if it is negatively biased. Table 4 shows that biases are small across models.

**[Table 4 around here]**

Finally, Table 5 summarizes average computational times. RE-ESF (A2), RE-ESF (A1), and RE-ESF are the first, second and third fastest, respectively. The computational efficiency of RE-ESF does not hold when either the sample size, $N$, or the number of SVCs, $K$, is large because RE-ESF requires optimizing the $2K$ parameters simultaneously. Base on Table 5, RE-ESF is slower than GWR if $N \geq 5000$. Still, RE-ESF (A1), whose coefficient estimates are as accurate as those for RE-ESF, is faster than GWR. Use of RE-ESF (A1), which allows spatial variation only for several focused coefficients, is a sensible option to reduce computational cost. Note that although ESF involves the computing slowest because of the eigenvector selection step, this step can be replaced with computationally more efficient approaches, such as lasso estimation (Seya et al., 2015).

**[Table 5 around here]**

## 7. An application to a land price analysis

This section empirically compares SVC models. Results show that ESF-based and RE-ESF-based SVC models are robust to multicollinearity, and they furnish reasonable SVC estimates for actual data.

### 7.1. Outline

This section presents an application of GWR, LCR-GWR, the ESF-based SVC model, and the RE-ESF-based SVC model to analyze land price and flood risk in Ibaraki prefecture, Japan. The western part of Ibaraki was seriously damaged by a river flood in September 2015 (see Figure 1). By December 21, 2015, 54 residences were totally destroyed, 3,752 suffered large-scale partial destruction, and 208 were partially destroyed, while about 10,390 people were in shelters at the peak of the disaster.

**[Figure 1 around here]**

Our goal here is to assess whether high risk areas were appropriately recognized as less attractive areas before the flood. To examine this concern, we analyze the relationship between flood hazards and land prices. Specifically, logged officially assessed land prices in 2015 (sample size: 647; see Figure 1 and Table 6) are described using the aforementioned SVC models. The response variables are flood depth (*Flood*), distance to the nearest railway station in km (*Station_D*), and railway distance between the nearest station and Tokyo station (*Tokyo_D*), which is located about 30 km from the southwestern border of Ibaraki. All of these variables measures are available from the National Land Numerical Information download service provided by the Ministry of Land, Infrastructure, Transport and Tourism (http://nlftp.mlit.go.jp/ksj-e/index.html).

VIFs of these variables are 1.09, 1.02 and 1.07, respectively. Thus, serious multicollinearity is not present among them. Since the main objective of this analysis is to compare the SVC models, including the GWR model, which loses degrees of freedom drastically as the number of explanatory variables increases (Griffith, 2008), we restricted the number of explanatory variables to three.

**[Table 6 around here]**

This empirical analysis is performed by employing R version 3.1.1 for computation purposes, and ArcGIS version 10.3 (http://www.esri.com/) for visualization. R and ArcGIS were executed on a 64 bit PC whose memory is 48 GB. The 'GWmodel' package in R was used to estimate GWR and LCR-GWR parameters.

*7.2. Results*

Hereafter, the vector of the spatially varying intercepts is denoted by $\boldsymbol{\beta}_0$ and those of the spatially varying coefficients for *Tokyo_D*, *Station_D*, and *Flood* are denoted by $\boldsymbol{\beta}_{\mathrm{Tk}}$, $\boldsymbol{\beta}_{\mathrm{St}}$, and $\boldsymbol{\beta}_{\mathrm{Fl}}$, respectively.

Table 7 summarizes the variance parameters ($\sigma^2_{k\,(\gamma)}$ and $\alpha_k$) estimated by RE-ESF. $\sigma^2_{k\,(\gamma)} = 0$ regarding $\boldsymbol{\beta}_{\mathrm{Tk}}$ shows that the impact of *Tokyo_D* is constant across the target area.[6] The positive $\sigma^2_{k(\gamma)}$ values for $\boldsymbol{\beta}_0$, $\boldsymbol{\beta}_{\mathrm{St}}$, and $\boldsymbol{\beta}_{\mathrm{Fl}}$ suggest that each has spatial variation.

**[Table 7 around here]**

The spatial smoothness (or scale) of $\boldsymbol{\beta}_{\mathrm{Fl}}$ is strongly intensified by a large $\alpha_k$ value. In contrast, the spatial smoothness of $\boldsymbol{\beta}_{\mathrm{St}}$, whose $\alpha_k$ equals zero, is not intensified. Although the bandwidths estimated by GWR and LCR-GWR (1.53 km and 2.77 km, respectively) suggest the existence of local spatial variations in each coefficient; based on the $\alpha_k$ values, bandwidths might actually differ across coefficients. More specifically, the bandwidths of $\boldsymbol{\beta}_{St}$, $\boldsymbol{\beta}_{Fl}$, and $\boldsymbol{\beta}_{Tk}$ are likely to be small, moderate and very large,[7] respectively.

---

[6] The variance becomes zero even when we apply RE-ESF (A1).

Figure 2 displays the boxplots of the estimated coefficients. While the boxplots of $\boldsymbol{\beta}_{St}$ are similar across the models, the variance of $\boldsymbol{\beta}_{Fl}$ is inflated in GWR, and those of $\boldsymbol{\beta}_{0}$ and $\boldsymbol{\beta}_{Tk}$ are highly inflated in GWR and LCR-GWR. For example, while logged land prices take values between 8.57 and 12.58, $\boldsymbol{\beta}_{0}$ estimated by GWR ranges between –5.76 and 26.15.

**[Figure 2 around here]**

The variance inflation might be because GWR and LCR-GWR rely on local estimations. Because *Tokyo_D* has a global map pattern, its variations tend to be small in each local subsample. As a result, GWR models might fail to differentiate influences from *Tokyo_D*, with small variations, and intercepts with no variation. Wheeler (2010) also reports a similar problem. Because ESF and RE-ESF consider all samples in their estimation, their coefficients are more stable, even if some of the covariates have global map patterns. Interestingly, the boxplots of the ESF coefficients are similar to those of the RE-ESF coefficients.

[7] The coefficients of GWR are constant when the bandwidth is extremely large.

Although the variance of $\boldsymbol{\beta}_{Fl}$ in GWR also is inflated, it is moderated for LCR-GWR. Effectiveness of the regularized GWR approach is verified. ESF and RE-ESF also provide stable coefficient estimates.

Table 8 summarizes correlation coefficients among SVCs. $\boldsymbol{\beta}_0$ and $\boldsymbol{\beta}_{Tk}$ have strong negative correlations for the GWR and LCR-GWR models. The greater variations of $\boldsymbol{\beta}_0$ and $\boldsymbol{\beta}_{Tk}$, portrayed in Figure 2, are attributable to their multicollinearity. By contrast, correlation coefficients for the ESF and RE-ESF models are reasonably small, and no serious multicollinearity was found. The result is consistent with a suggestion by Griffith (2008) that the ESF-based specification is robust to multicollinearity.

**[Table 8 around here]**

Figure 3 plots the estimated coefficients. In each model, the estimated $\boldsymbol{\beta}_0$ demonstrates greater land prices in the nearby Tokyo area and around Mito city, which is the prefectural capital. The spatial distributions of $\boldsymbol{\beta}_{St}$ suggest that land prices decline rapidly as distance to the nearest station increases in nearby station areas, whereas this reduction is moderated in suburban areas. The estimated $\boldsymbol{\beta}_0$ and $\boldsymbol{\beta}_{St}$ are similar across models.

**[Figure 3 around here]**

Consistent with the expected negative sign of $\boldsymbol{\beta}_{Tk}$, 643/648 of its elements for ESF, and all of its elements for RE-ESF are negative. In contrast, 465/648 and 10/648 elements are positive in the GWR and LCR-GWR models, respectively, probably because of the variance inflation previously discussed. Another notable difference is that RE-ESF $\boldsymbol{\beta}_{Tk}$ estimates have no spatial variation (i.e., $\sigma^2_{k(\gamma)} = 0$), whereas the other $\boldsymbol{\beta}_{Tk}$ estimates that have significant spatial variation.

The elements of $\boldsymbol{\beta}_{\mathrm{Fl}}$ are negative if flood-prone areas have lower land prices. $\boldsymbol{\beta}_{\mathrm{Fl}}$ obtained from RE-ESF displays a smoother map pattern than for the other models because of the large $\alpha_k$ value (3.02). The $\boldsymbol{\beta}_{\mathrm{Fl}}$ for RE-ESF is negative around Mito, where high risk areas are appropriately recognized as less attractive. In contrast, $\boldsymbol{\beta}_{\mathrm{Fl}}$ is positive in the western area, including the area flooded in September 2015. In other words, high risk areas are recognized as attractive areas. This result implies that benefits of rivers (e.g., natural environment, landscape) are emphasized more than flood risk. This situation may have increased the resulting damage from the 2015 flood. In contrast, the $\boldsymbol{\beta}_{\mathrm{Fl}}$ estimated by the other models takes both positive and negative values in the flooded area.

Finally, Table 9 summarizes the computational times. For reference, the computational time of RE-ESF (A1) also is calculated and included. This table shows that RE-ESF is computationally more efficient than LCR-GWR and ESF in the case of $N = 647$ and three covariates. Furthermore, computation of estimates for RE-ESF (A1) is more than three times faster than for GWR in this case. Of note is that GWR calculations are faster than RE-ESF(A1) if sample size is large. This timing difference is because of the requirement of an eigen-decomposition.

**[Table 9 around here]**

In summary, we empirically verified that each SVC model can provide different results, and that the estimates of RE-ESF seem reliable (i.e., interpretable and displaying smaller variance).

## 8. Concluding remarks

This study proposes an RE-ESF-based SVC model whose coefficients are interpretable based on the MC. A simulation analysis and an empirical analysis involving land prices suggest advantages of our model in terms of estimation accuracy, computational time, and interpretability of coefficient estimates.

Unlike GWR models and the typical B-SVC model, RE-ESF estimates the smoothness of each SVC in a computationally efficient manner. Although coefficient smoothness parameters also can be introduced into the B-SVC model, their estimation is computationally prohibitive. Thus, our approach is useful as a flexible and relatively simple procedure. Meanwhile, computationally efficient and flexible alternatives, including the integrated nested Laplace approximation (INLA: Rue et al., 2009)-based SVC model (Congdon, 2014), have been proposed recently. Therefore, the comparison of our model with these is therefore an important future research topic.

Another remaining issue is to compare our model with other SVC models from the viewpoint of statistical inference for the $\boldsymbol{\beta}_k$s. We also need to examine the validity of our model in cases with many covariates for which multicollinearity among SVCs can be serious. Furthermore, extension of our model to a wide range of applications would be an interesting next step. These extensions might include change of support problems (e.g., Murakami and Tsutsumi, 2012; 2015), interaction data modeling (Chun and Griffith, 2011), non-Gaussian data modeling (Fotheringham et al., 2002; Nakaya et al., 2005; Griffith, 2011), spatiotemporal data modeling (Fotheringham et al., 2015; Huang et al., 2010; Griffith, 2012).

## Acknowledgements

This work was supported by Grants-in-Aid for Scientific Research from JSPS (Research Project Number: 15H04054). We thank Professor Morito Tsutsumi (University of Tsukuba) and Professor Takahisa Yokoi (Tohoku University) for their suggestive advices.

**Table 1.** RMSEs of the estimated coefficients ($N = 50$)

| Coef. | $w_s$ | $r_1$ | $r_2$ | GWR | LCR-GWR | ESF | RE-ESF | RE-ESF(A1) | RE-ESF(A2) |
|---|---|---|---|---|---|---|---|---|---|
| $\boldsymbol{\beta}_1$ | 0.0 | 0.5 | 1.0 | 2.32 | 2.32 | 2.58 | **2.23** | **2.25** | 2.58 |
| | | 1.0 | 0.5 | 2.27 | 2.28 | 2.57 | **2.22** | **2.25** | 2.58 |
| | | 1.0 | 1.0 | 2.29 | 2.29 | 2.56 | **2.23** | **2.25** | 2.55 |
| | | 1.0 | 2.0 | 2.33 | 2.33 | 2.58 | **2.25** | **2.27** | 2.59 |
| | | 2.0 | 1.0 | **2.43** | **2.44** | 2.86 | 2.44 | 2.46 | 2.66 |
| | 0.4 | 0.5 | 1.0 | 2.45 | 2.45 | 3.03 | 2.40 | **2.40** | 2.70 |
| | | 1.0 | 0.5 | 2.40 | 2.40 | 2.94 | **2.36** | **2.38** | 2.73 |
| | | 1.0 | 1.0 | **2.39** | 2.39 | 3.06 | **2.39** | 2.39 | 2.67 |
| | | 1.0 | 2.0 | 2.47 | 2.44 | 3.01 | **2.38** | **2.40** | 2.74 |
| | | 2.0 | 1.0 | **2.56** | **2.56** | 3.24 | 2.59 | 2.60 | 2.80 |
| | 0.8 | 0.5 | 1.0 | 2.83 | 2.78 | 3.78 | **2.63** | **2.64** | 3.05 |
| | | 1.0 | 0.5 | 2.74 | 2.69 | 3.54 | **2.55** | **2.59** | 3.01 |
| | | 1.0 | 1.0 | 2.68 | 2.68 | 3.75 | **2.55** | **2.58** | 2.98 |
| | | 1.0 | 2.0 | 2.62 | 2.62 | 3.67 | **2.50** | **2.53** | 2.98 |
| | | 2.0 | 1.0 | 2.71 | **2.68** | 3.83 | **2.68** | 2.71 | 2.96 |
| $\boldsymbol{\beta}_2$ | 0.0 | 0.5 | 1.0 | 1.13 | 1.12 | 1.79 | **1.11** | 1.14 | **1.12** |
| | | 1.0 | 0.5 | 1.15 | 1.14 | 1.82 | **1.11** | 1.14 | **1.11** |
| | | 1.0 | 1.0 | 1.10 | **1.08** | 1.80 | **1.08** | 1.09 | 1.10 |
| | | 1.0 | 2.0 | 1.16 | 1.15 | 1.84 | **1.10** | 1.12 | **1.11** |
| | | 2.0 | 1.0 | **1.09** | **1.09** | 1.86 | 1.12 | 1.14 | 1.12 |
| | 0.4 | 0.5 | 1.0 | 1.25 | 1.24 | 2.36 | **1.23** | 1.26 | **1.18** |
| | | 1.0 | 0.5 | 1.26 | 1.26 | 2.22 | **1.25** | 1.27 | **1.19** |
| | | 1.0 | 1.0 | 1.27 | 1.26 | 2.25 | **1.25** | 1.27 | **1.19** |
| | | 1.0 | 2.0 | 1.30 | **1.27** | 2.29 | 1.31 | 1.36 | **1.20** |
| | | 2.0 | 1.0 | **1.22** | 1.22 | 2.60 | 1.26 | 1.29 | **1.18** |
| | 0.8 | 0.5 | 1.0 | 1.57 | 1.53 | 2.97 | **1.48** | 1.52 | **1.42** |
| | | 1.0 | 0.5 | 1.51 | 1.46 | 2.81 | **1.33** | 1.37 | 1.41 |
| | | 1.0 | 1.0 | 1.52 | 1.49 | 2.99 | **1.41** | 1.45 | **1.40** |
| | | 1.0 | 2.0 | 1.40 | 1.40 | 2.71 | **1.35** | **1.38** | 1.40 |
| | | 2.0 | 1.0 | 1.40 | **1.36** | 3.07 | 1.37 | 1.39 | **1.30** |

Note: $w_s$ intensifies the spatial dependence in each coefficient; $r_1$ and $r_2$ are the bandwidths in Eq. (30) for $\boldsymbol{\beta}_1$ and $\boldsymbol{\beta}_2$, respectively. Dark gray denotes the minimum RMSE in each case, and light gray denotes the second minimum RMSE.

**Table 2.** RMSEs of the estimated coefficients ($N = 150$)

| Coef. | $w_s$ | $r_1$ | $r_2$ | GWR | LCR-GWR | ESF | RE-ESF | RE-ESF (A1) | RE-ESF (A2) |
|---|---|---|---|---|---|---|---|---|---|
| $\beta_1$ | 0.0 | 0.5 | 1.0 | 1.87 | 1.87 | 1.99 | **1.77** | **1.78** | 1.99 |
| | | 1.0 | 0.5 | 1.61 | 1.62 | 1.86 | **1.57** | **1.57** | 1.78 |
| | | 1.0 | 1.0 | 1.66 | 1.67 | 1.89 | **1.60** | **1.60** | 1.76 |
| | | 1.0 | 2.0 | 1.60 | 1.61 | 1.86 | **1.54** | **1.54** | 1.76 |
| | | 2.0 | 1.0 | **1.75** | 1.76 | 1.98 | **1.75** | 1.76 | 1.91 |
| | 0.4 | 0.5 | 1.0 | 2.04 | 2.04 | 2.33 | **1.92** | **1.93** | 2.23 |
| | | 1.0 | 0.5 | 1.79 | 1.79 | 2.22 | **1.73** | **1.73** | 2.05 |
| | | 1.0 | 1.0 | 1.76 | 1.76 | 2.18 | **1.68** | **1.69** | 2.00 |
| | | 1.0 | 2.0 | 1.83 | 1.83 | 2.15 | **1.76** | **1.77** | 2.04 |
| | | 2.0 | 1.0 | 1.85 | 1.85 | 2.27 | **1.81** | **1.81** | 2.03 |
| | 0.8 | 0.5 | 1.0 | 2.44 | 2.41 | 3.16 | **2.11** | **2.13** | 2.75 |
| | | 1.0 | 0.5 | 2.15 | 2.14 | 2.92 | **1.85** | **1.89** | 2.50 |
| | | 1.0 | 1.0 | 2.11 | 2.10 | 3.00 | **1.85** | **1.87** | 2.53 |
| | | 1.0 | 2.0 | 2.16 | 2.15 | 2.87 | **1.93** | **1.93** | 2.45 |
| | | 2.0 | 1.0 | 2.15 | 2.15 | 3.00 | **1.98** | **2.00** | 2.51 |
| $\beta_2$ | 0.0 | 0.5 | 1.0 | 0.88 | 0.87 | 1.32 | **0.80** | 0.81 | **0.78** |
| | | 1.0 | 0.5 | 0.90 | 0.90 | 1.29 | **0.87** | 0.87 | **0.84** |
| | | 1.0 | 1.0 | 0.84 | 0.84 | 1.29 | **0.78** | 0.79 | **0.76** |
| | | 1.0 | 2.0 | 0.88 | 0.88 | 1.28 | **0.80** | 0.80 | **0.79** |
| | | 2.0 | 1.0 | 0.81 | 0.81 | 1.28 | **0.80** | 0.80 | **0.79** |
| | 0.4 | 0.5 | 1.0 | 1.05 | 1.04 | 1.70 | **0.95** | 0.96 | **0.91** |
| | | 1.0 | 0.5 | 1.06 | 1.05 | 1.68 | **0.98** | 0.98 | **0.94** |
| | | 1.0 | 1.0 | 0.98 | 0.98 | 1.65 | 0.94 | **0.93** | **0.86** |
| | | 1.0 | 2.0 | 0.98 | 0.98 | 1.62 | **0.91** | 0.92 | **0.87** |
| | | 2.0 | 1.0 | 0.94 | 0.93 | 1.61 | **0.93** | 0.93 | **0.88** |
| | 0.8 | 0.5 | 1.0 | 1.49 | 1.44 | 2.58 | **1.20** | **1.23** | 1.24 |
| | | 1.0 | 0.5 | 1.45 | 1.43 | 2.35 | 1.18 | **1.18** | 1.23 |
| | | 1.0 | 1.0 | 1.37 | 1.36 | 2.40 | **1.12** | **1.15** | 1.21 |
| | | 1.0 | 2.0 | 1.34 | 1.32 | 2.24 | **1.09** | 1.12 | **1.11** |
| | | 2.0 | 1.0 | 1.27 | 1.27 | 2.48 | **1.10** | **1.12** | 1.16 |

Note: See Table 1.

**Table 3.** RMSEs of the estimated coefficients ($N$ = 400)

| Coef. | $w_s$ | $r_1$ | $r_2$ | GWR | LCR-GWR | ESF | RE-ESF | RE-ESF (A1) | RE-ESF (A2) |
|---|---|---|---|---|---|---|---|---|---|
| $\beta_1$ | 0.0 | 0.5 | 1.0 | 1.44 | 1.45 | 1.56 | **1.36** | **1.36** | 1.55 |
| | | 1.0 | 0.5 | 1.20 | 1.21 | 1.33 | **1.13** | **1.13** | 1.24 |
| | | 1.0 | 1.0 | 1.23 | 1.24 | 1.35 | **1.17** | **1.16** | 1.30 |
| | | 1.0 | 2.0 | 1.20 | 1.21 | 1.30 | **1.13** | **1.12** | 1.24 |
| | | 2.0 | 1.0 | 1.23 | 1.24 | 1.41 | **1.20** | **1.19** | 1.25 |
| | 0.4 | 0.5 | 1.0 | 1.63 | 1.61 | 1.83 | **1.49** | **1.49** | 1.76 |
| | | 1.0 | 0.5 | 1.34 | 1.34 | 1.69 | **1.25** | **1.25** | 1.42 |
| | | 1.0 | 1.0 | 1.34 | 1.35 | 1.58 | **1.25** | **1.25** | 1.45 |
| | | 1.0 | 2.0 | 1.33 | 1.33 | 1.61 | **1.23** | **1.22** | 1.41 |
| | | 2.0 | 1.0 | 1.37 | 1.37 | 1.65 | **1.31** | **1.31** | 1.42 |
| | 0.8 | 0.5 | 1.0 | 2.25 | 2.15 | 2.67 | **1.74** | **1.75** | 2.31 |
| | | 1.0 | 0.5 | 1.86 | 1.83 | 2.39 | **1.46** | **1.46** | 1.93 |
| | | 1.0 | 1.0 | 1.80 | 1.78 | 2.37 | **1.45** | **1.45** | 1.96 |
| | | 1.0 | 2.0 | 1.84 | 1.81 | 2.34 | **1.48** | **1.48** | 1.95 |
| | | 2.0 | 1.0 | 1.72 | 1.71 | 2.30 | **1.46** | **1.46** | 1.90 |
| $\beta_2$ | 0.0 | 0.5 | 1.0 | 0.75 | 0.74 | 0.96 | **0.58** | **0.59** | 0.59 |
| | | 1.0 | 0.5 | 0.72 | 0.72 | 0.97 | **0.66** | 0.66 | **0.65** |
| | | 1.0 | 1.0 | 0.67 | 0.67 | 0.90 | **0.56** | 0.57 | **0.56** |
| | | 1.0 | 2.0 | 0.65 | 0.66 | 0.92 | **0.55** | **0.55** | 0.55 |
| | | 2.0 | 1.0 | 0.61 | 0.61 | 0.91 | **0.55** | 0.56 | **0.55** |
| | 0.4 | 0.5 | 1.0 | 0.92 | 0.90 | 1.30 | **0.71** | 0.71 | **0.68** |
| | | 1.0 | 0.5 | 0.87 | 0.86 | 1.31 | **0.77** | 0.77 | **0.74** |
| | | 1.0 | 1.0 | 0.82 | 0.82 | 1.22 | **0.68** | 0.69 | **0.66** |
| | | 1.0 | 2.0 | 0.81 | 0.81 | 1.24 | **0.65** | 0.66 | **0.64** |
| | | 2.0 | 1.0 | 0.79 | 0.79 | 1.25 | **0.69** | 0.70 | **0.67** |
| | 0.8 | 0.5 | 1.0 | 1.51 | 1.43 | 2.10 | **0.99** | **0.99** | 1.03 |
| | | 1.0 | 0.5 | 1.39 | 1.37 | 1.95 | **0.96** | **0.96** | 1.01 |
| | | 1.0 | 1.0 | 1.31 | 1.29 | 1.88 | **0.92** | **0.92** | 0.93 |
| | | 1.0 | 2.0 | 1.34 | 1.32 | 1.88 | **0.91** | **0.93** | 0.93 |
| | | 2.0 | 1.0 | 1.28 | 1.26 | 1.87 | **0.92** | **0.93** | 0.95 |

Note: See Table 1.

**Table 4.** Bias of the estimated coefficients ($r_1$ = 2; $r_2$ = 1): $w_s$ intensifies the spatial dependence in each coefficient.

| $N$ | Coef. | $w_s$ | GWR | LCR-GWR | ESF | RE-ESF | RE-ESF (A1) | RE-ESF (A2) |
|---|---|---|---|---|---|---|---|---|
| 50 | $\beta_1$ | 0.0 | **0.05** | 0.05 | –0.06 | 0.06 | 0.08 | **0.01** |
| | | 0.4 | –0.07 | –0.07 | –0.08 | **0.03** | **0.02** | –0.09 |
| | | 0.8 | –0.08 | –0.08 | **0.00** | **0.01** | –0.06 | –0.08 |
| | $\beta_2$ | 0.0 | –0.04 | –0.04 | –0.05 | **–0.04** | 0.08 | **0.01** |
| | | 0.4 | 0.07 | **0.07** | **–0.05** | 0.17 | 0.22 | 0.13 |
| | | 0.8 | **0.03** | **0.03** | –0.8 | 0.14 | 0.22 | 0.17 |
| 150 | $\beta_1$ | 0.0 | –0.02 | –0.02 | **–0.02** | **0.02** | 0.06 | –0.08 |
| | | 0.4 | **–0.01** | **–0.01** | –0.05 | –0.04 | –0.06 | 0.03 |
| | | 0.8 | **0.00** | **0.00** | 0.10 | 0.17 | 0.21 | 0.12 |
| | $\beta_2$ | 0.0 | 0.02 | **0.02** | **0.01** | 0.04 | 0.03 | 0.05 |
| | | 0.4 | **0.01** | **0.01** | –0.01 | –0.06 | –0.09 | –0.12 |
| | | 0.8 | –0.02 | –0.02 | 0.03 | **0.00** | **–0.01** | –0.01 |
| 400 | $\beta_1$ | 0.0 | 0.00 | **0.00** | –0.07 | –0.04 | –0.06 | **0.00** |
| | | 0.4 | **0.01** | **0.01** | –0.08 | –0.02 | –0.07 | 0.05 |
| | | 0.8 | **0.00** | 0.02 | **–0.02** | 0.07 | 0.10 | 0.17 |
| | $\beta_2$ | 0.0 | **0.00** | **0.00** | 0.02 | –0.01 | 0.00 | –0.02 |
| | | 0.4 | **0.00** | **0.00** | –0.02 | –0.04 | –0.08 | –0.01 |
| | | 0.8 | 0.01 | 0.01 | –0.03 | **–0.01** | **0.01** | 0.01 |

Note: Dark gray denotes the minimum bias in each case and light gray denotes the second minimum bias.

**Table 5.** Mean computational time in seconds ($r_1 = 2$; $r_2 = 1$).

| *N* | GWR | LCR-GWR | ESF | RE-ESF | RE-ESF (A1) | RE-ESF (A2) |
|---|---|---|---|---|---|---|
| 50 | 0.30 | 0.80 | 0.82 | 0.20 | 0.10 | 0.10 |
| 150 | 2.05 | 3.88 | 5.15 | 0.79 | 0.39 | 0.35 |
| 400 | 13.18 | 20.40 | 32.39 | 5.04 | 2.49 | 2.12 |
| 1,000 | 79.00 | 115.21 | 275.48 | 24.40 | 15.81 | 12.78 |
| 2,000 | 326.42 | 495.61 | 2056.01 | 75.33 | 122.06 | 65.96 |
| 5,000 | 1984.99 | 3465.37 | 56324.66 | 2241.90 | 1110.97 | 883.26 |

Note: Because computational times were very similar across iterations, regarding cases with $N$ = {1,000, 2,000, 3000}, we performed five replicates, and averaged the resulting five computational times.

**Table 6.** Summary statistics for land prices (JPY/$m^2$).

| Statistics | Value |
|---|---|
| Mean | 35.68 |
| Median | 29.50 |
| Standard error | 27.14 |
| Maximum | 290.00 |
| Minimum | 5.28 |
| Sample size | 647 |

**Table 7.** Estimates of the variance parameters in RE-ESF: $\sigma^2_{k(\gamma)}$ controls the variance of each coefficient, and $\alpha_k$ controls the spatial scale of their variations.

| | $\boldsymbol{\beta}_0$ | $\boldsymbol{\beta}_{Tk}$ | $\boldsymbol{\beta}_{St}$ | $\boldsymbol{\beta}_{Fl}$ |
|---|---|---|---|---|
| $\sigma^2_{k(\gamma)}$ | 1.71 | 0.00 | 0.35 | 0.61 |
| $\alpha_k$ | 0.27 | N.A.[1)] | 0.00 | 3.02 |

[1)] Because $\boldsymbol{\beta}_{St}$ lacks spatial variation (i.e., $\sigma^2_{k(\gamma)} = 0.00$), $\alpha_k$ for $\boldsymbol{\beta}_{St}$ is undefined.

**Table 8.** Correlation coefficients among SVCs.

GWR

| | $\beta_0$ | $\beta_{Tk}$ | $\beta_{St}$ | $\beta_{Fl}$ |
|---|---|---|---|---|
| $\beta_0$ | | -0.87 | -0.27 | 0.03 |
| $\beta_{Tk}$ | | | 0.23 | -0.03 |
| $\beta_{St}$ | | | | -0.08 |

LCR-GWR

| | $\beta_0$ | $\beta_{Tk}$ | $\beta_{St}$ | $\beta_{Fl}$ |
|---|---|---|---|---|
| $\beta_0$ | | -0.74 | -0.45 | -0.18 |
| $\beta_{Tk}$ | | | 0.24 | 0.26 |
| $\beta_{St}$ | | | | 0.34 |

ESF

| | $\beta_0$ | $\beta_{Tk}$ | $\beta_{St}$ | $\beta_{Fl}$ |
|---|---|---|---|---|
| $\beta_0$ | | 0.01 | -0.44 | 0.08 |
| $\beta_{Tk}$ | | | -0.24 | -0.23 |
| $\beta_{St}$ | | | | -0.01 |

RE-ESF

| | $\beta_0$ | $\beta_{Tk}$ | $\beta_{St}$ | $\beta_{Fl}$ |
|---|---|---|---|---|
| $\beta_0$ | | NA [1)] | -0.41 | -0.29 |
| $\beta_{Tk}$ | | | NA | NA |
| $\beta_{St}$ | | | | -0.10 |

[1)] Regarding RE-ESF, correlation coefficients between $\beta_{Tk}$ and other coefficients cannot be calculated because it lacks spatial variations (i.e., $\sigma^2_{k(\gamma)} = 0$).

**Table 9.** Computational time in seconds.

| GWR | LCR-GWR | ESF | RE-ESF | RE-ESF (A1) |
|---|---|---|---|---|
| 36.2 | 62.3 | 107 | 51.7 | 11.8 |

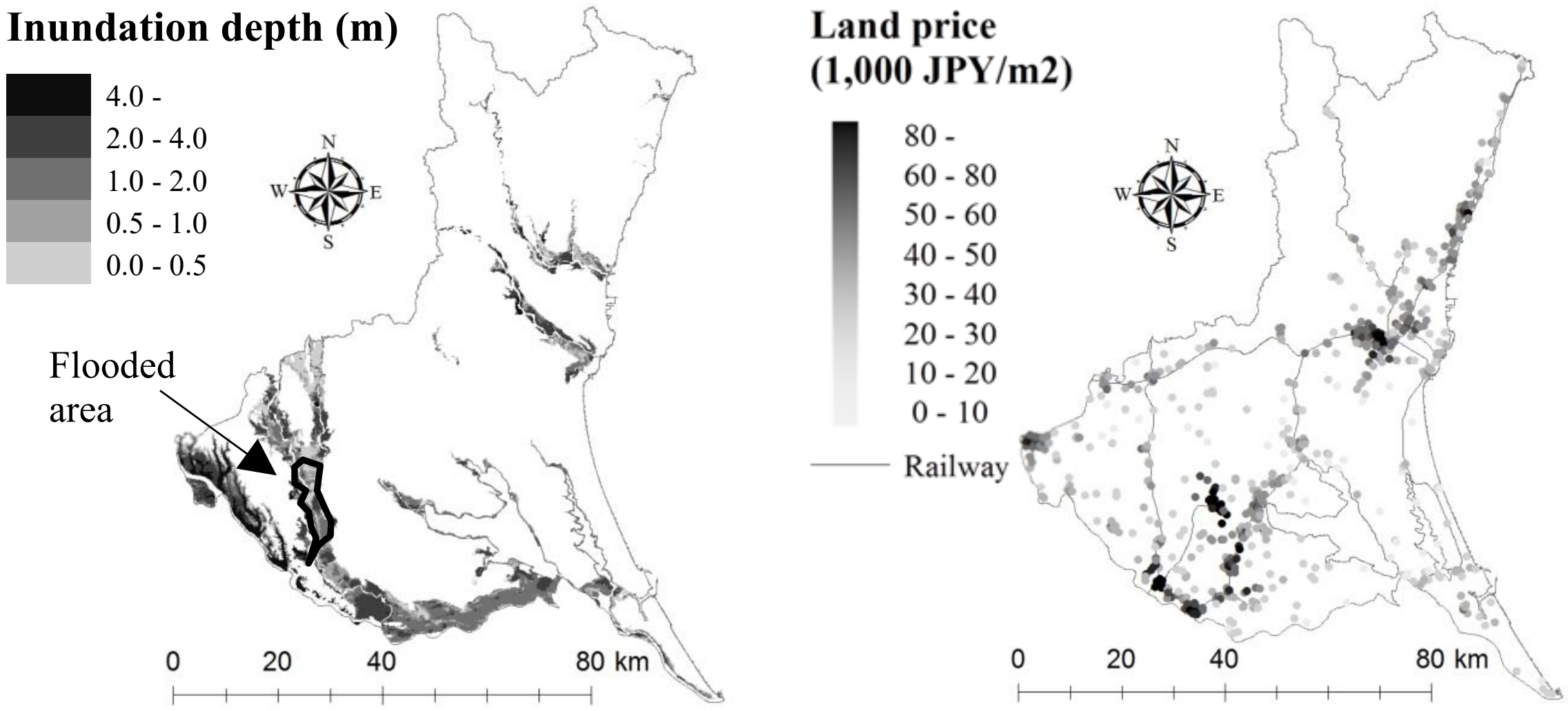

**Figure 1.** Anticipated flood depth (left) and officially assessed land prices in 2015 (right) in the Ibaraki prefecture.

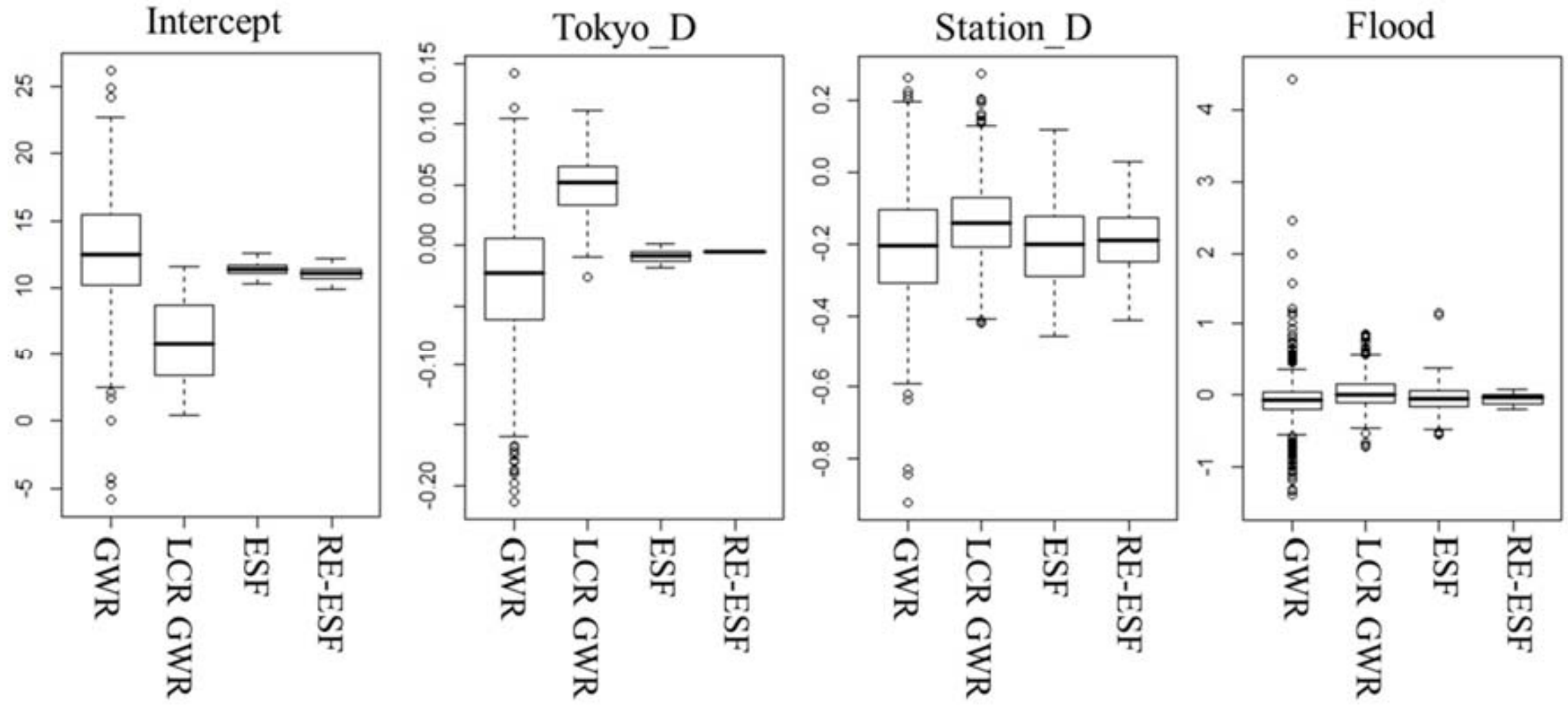

**Figure 2.** Boxplots of the estimated spatially varying coefficients.

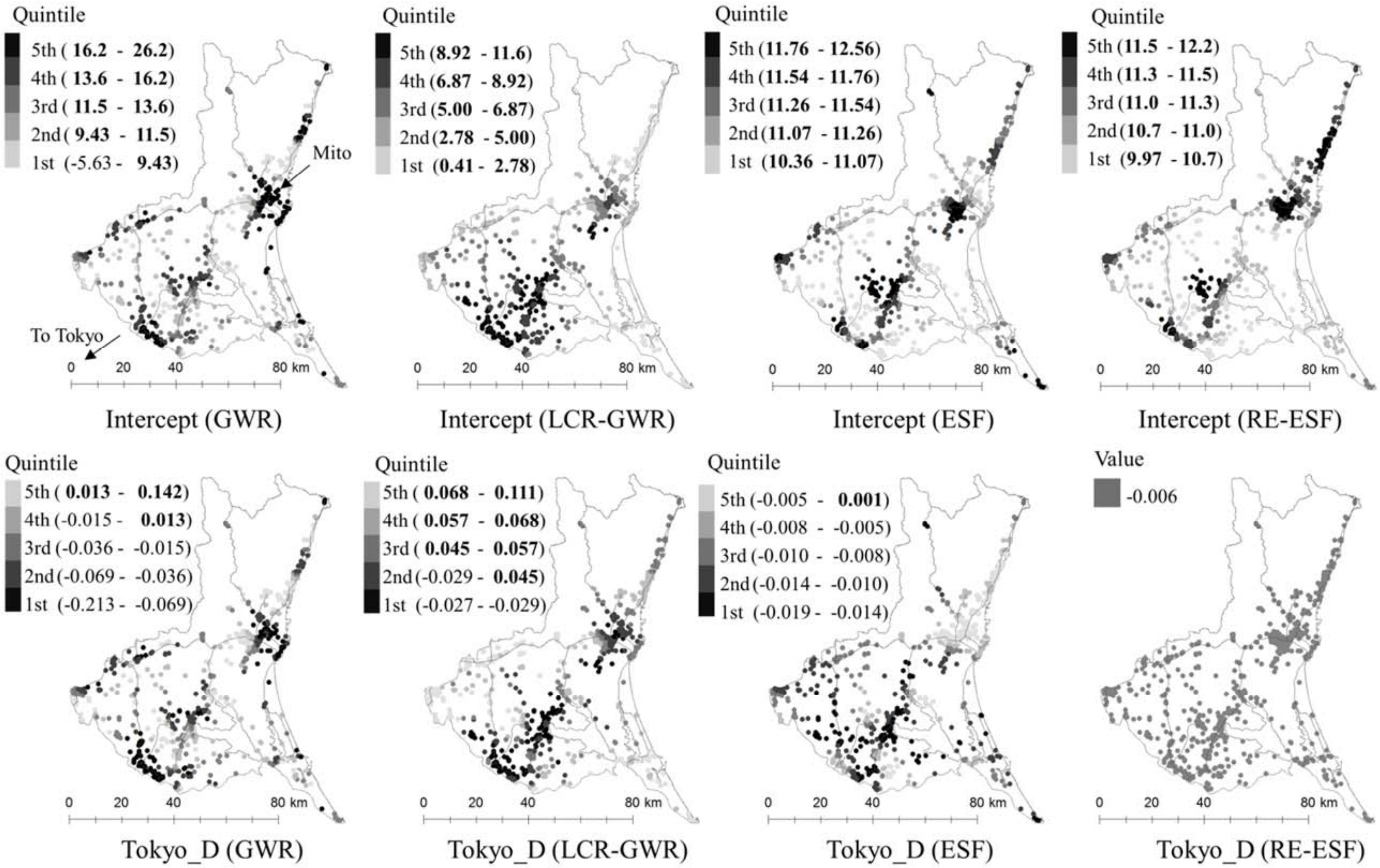

**Figure 3.** Spatial plots of the estimated coefficients. In each legend, positive values are denoted by bold text.

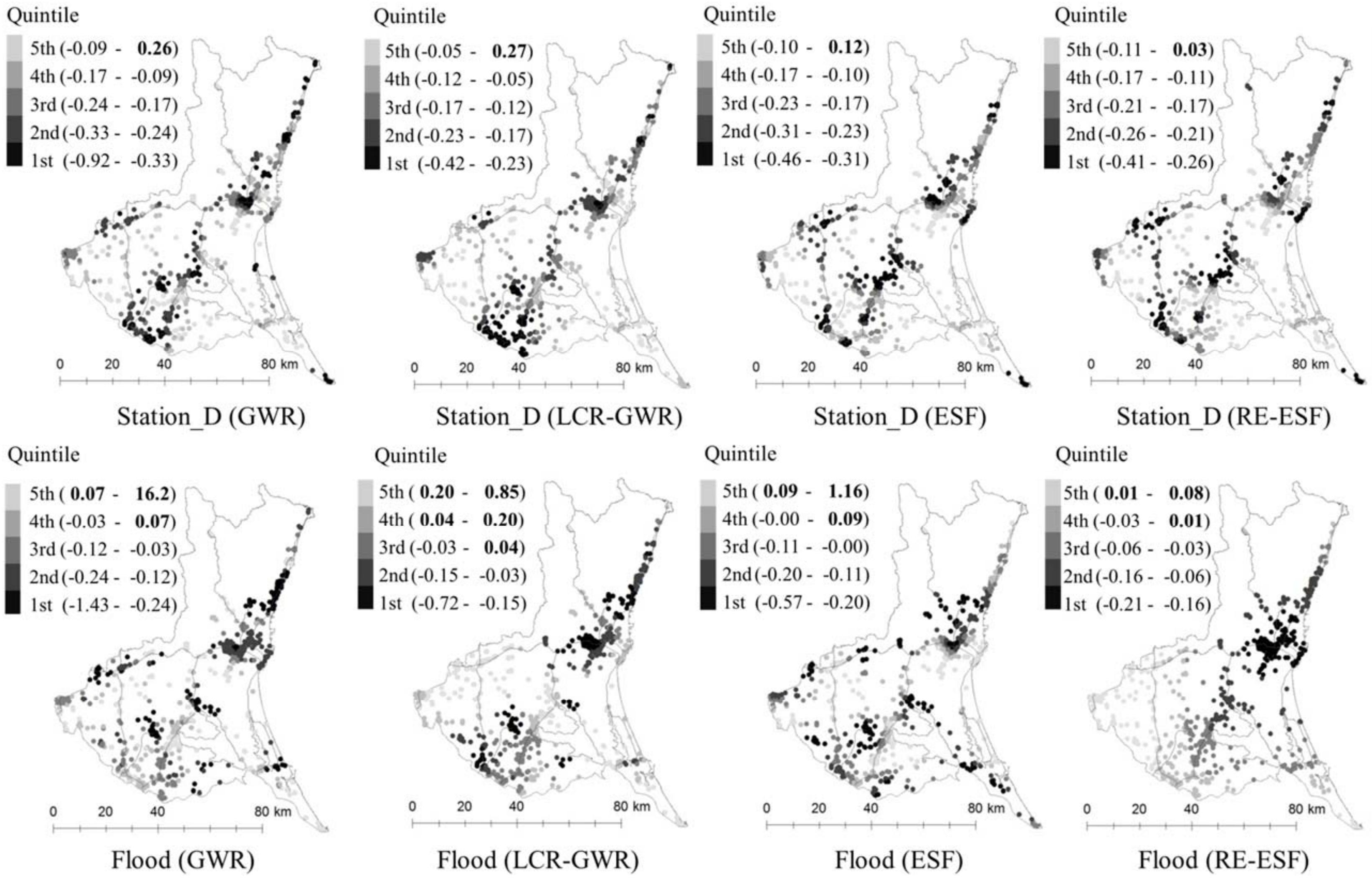

**Figure 3.** Spatial plots of the estimated coefficients (continued). In each legend, positive values are denoted by bold text.